\documentclass[aps,prb,twocolumn,showpacs]{revtex4}

\usepackage{amsmath}
\usepackage{graphicx}
\usepackage{setspace}
\usepackage{bm}
\usepackage{dcolumn}
\usepackage{ulem}

\begin{document}

\title{Determination of critical current density from arbitrary flux relaxation process}

\author{Rongchao Ma} 
\affiliation{Department of Physics, University of Alberta, Edmonton, AB, Canada T6G 2G7}
\email{marongchao@yahoo.com}

\date{\today}

\begin{abstract}
The current-carrying ability of a type-II superconductor is generally represented by its critical current density. This can be determined by measuring a flux relaxation process starting with a testing current density that is greater than or equal to the critical value. Here we show that a flux relaxation process starting with an intermediate current density can be converted into a process starting with the critical current density by introducing a virtual time interval. Therefore, one may calculate the critical current density from the flux relaxation process starting with a current density below the critical value. The exact solutions of the time dependence of current density in the flux relaxation process were also discussed.
\end{abstract}


\maketitle

\section{Introduction}

Critical current density is the maximum current density that can be carried by a type-II superconductor. Its value is mainly determined by the vortex pinning potential. One can determine the pinning potential from a flux relaxation process, which is the phenomenon that quantized vortices jump between adjacent pinning centers spontaneously due to various reasons \cite{Landau,Koren,Nicodemi,Hoekstra}. The flux relaxation phenomenon can be described by the Arrhenius equation in case of thermal activation. This suggests that a flux relaxation process has a strong dependence on temperature and vortex activation energy.\cite{Tinkham} The vortex activation energy is a decreasing function of current density because it can be reduced by the Lorentz force of the current density. A number of current dependent vortex activation energies \cite{Feigel'man1,Feigel'man2,Zeldov1,Zeldov2,Anderson,Ma1} were proposed on the basis of different physical considerations. Using these vortex activation energies, one can obtain the corresponding time evolution equations of the current density.\cite{Tinkham}

However, the time evolution equations of the current density give the critical current density at zero time. This means that, to find out the critical current density of a type-II superconductor, one has to apply a current density that is greater than or equal to the critical value. But at lower temperatures, the critical current density may be large \cite{Larbalestier} and the measurements are technically difficult. In fact, many measurements were carried out with an initial current density well below the critical value.\cite{Miclea,Miu,Harada,Lessure,Chikumoto} Thus, it is desirable to find a general formula which can be used to calculate the critical current density by measuring a flux relaxation process with an initial current density below the critical value.

For the above reasons, one should consider the vortex pinning potential, which plays a dominant role in determining the critical current density of a type-II superconductor. It is known that, at a constant temperature, the vortex pinning potential does not change in a flux relaxation process although the vortex activation energy does. This indicates that the effects of the vortex pinning potential are embodied in each vortex jump. Therefore, a vortex system may carry the information about the critical current density even with a testing current density well below the critical value. An early study has shown that the maximum internal field in a vortex penetration or flux relaxation process can be calculated by introducing some time parameters.\cite{Ma2} In case of current density, it should be possible to construct a theoretical model with which one can predict the critical current density by measuring a flux relaxation process with an arbitrary initial current density.

In this work, we first studied the time dependence of the vortex activation energies in two thought flux relaxation processes. The mathematical expression and physical meaning of a virtual time interval are discussed. Next, we used the inverse-power activation energy and logarithmic activation energy to construct the time evolution equations of the current density. Finally, we discuss the exact solutions of the time dependence of current density with linear activation energy and logarithmic activation energy. Let us now derive the generalized expression for the time dependence of the vortex activation energy.

\section{Time dependence of activation energy}

In the flux relaxation process, the current density $j$ can reduce the vortex activation energy $U_a$ by its Lorentz force and therefore increase the vortex hopping rate. This indicates that $U_a$ is a decreasing function of $j$. Because $j$ is a decreasing function of the time $t$, $U_a$ is an increasing function of $t$.

The rate of change of the current density, $dj/dt$, is proportional to the vortex hopping rate. Furthermore, the current density in a flux relaxation process is a decreasing function of the time, i.e., $dj/dt<0$. According to the Arrhenius equation, therefore, we have the following equation \cite{Geshkenbein,Beasley,Ma2}
\begin{equation}
\label{UaDiff}
\frac{dj}{dt} = - C e^{-U_a/kT},
\end{equation} 
where $C$ is a positive proportional constant.

The exact solution of Eq.(\ref{UaDiff}) is currently unavailable, but we can find an approximate solution to it. Consider a superconductor in which an arbitrary current density starts to decay at zero time ($t=0$). The initial vortex activation energy is $U_i$. As the time increases to $t$, the vortex activation energy increases to $U_a$. Rewrite Eq.(\ref{UaDiff}) and integrate it on both sides

\[ \int_{U_i}^{U_a} e^{U_a/kT}dU_a = - \int_0^t C\frac{dU_a}{dj} dt. \]

With logarithmic accuracy \cite{Geshkenbein}, we obtain the following equation
\begin{equation}
\label{UaTime1}
U_a(t) = kT ln\left( e^{U_i/kT} + \frac{t}{\tau} \right),
\end{equation} 
where $\tau = - kT/[C(dU_a/dj)]$ is a short time scale parameter \cite{Feigel'man2}. Here we included a negative sign to ensure that $\tau$ is a positive number because in a flux relaxation process the activation energy $U_a$ is a decreasing function of the current density $j$, i.e., $dU_a/dj<0$.

Eq.(\ref{UaTime1}) is the general time evolution equation of the vortex activation energy in a flux relaxation process starting with an arbitrary current density. If the flux relaxation process starts with the critical current density $j_c$, then the initial vortex activation energy $U_i$ is zero, i.e., $U_i|_{j=j_c}=0$. Therefore, Eq.(\ref{UaTime1}) reduces to 
\begin{equation}
\label{UaTime2} 
U_{ac}(t) = kT ln\left(1 + \frac{t}{\tau} \right).
\end{equation}

Eq.(\ref{UaTime2}) is the well-known time dependent vortex activation energy used in earlier theories \cite{Geshkenbein}, from which the current decay models were derived.

Because our purpose is to calculate the critical current density $j_c$ from a flux relaxation process with an initial current density $j_i$ below the critical value, we need to consider the possibility of converting Eq.(\ref{UaTime1}) into the form of Eq.(\ref{UaTime2}). To do this, let us first study the following two thought flux relaxation processes:

Case A: At temperature $T$, a vortex system in a superconductor starts to decay with an initial current density $j_i$ below the critical value where the vortex activation energy is $U_i$. The time dependence of the vortex activation energy is then governed by Eq.(\ref{UaTime1}), that is, $U_a(t) = kT ln (e^{U_i/kT} + t/\tau )$.

Case B: At temperature $T$, consider the same superconductor used in Case A. But a vortex system starts to decay with the critical current density $j_c$, where the initial vortex activation energy is $0$. After a time interval $t_i$, the vortex system and current density reduces to exactly the same as that in Case A, where the vortex activation energy is $U_i$. If we now restart counting the time $t$, then the flux relaxation process should be governed by Eq.(\ref{UaTime2}) with an argument $t_i + t$, that is, $U_{ac}(t_i+t) = kT ln [1 + (t_i+t)/\tau ]$.

Since Case A and Case B describe the same physical process at time $t$, they must obey the same physical law. Therefore, the vortex activation energies should be equal in quantity for all time $t$, that is,
\begin{equation}
\label{UU}
U_a(t) = U_{ac}(t_i+t).
\end{equation}

Substituting the exact expressions of $U_a(t)$ and $U_{ac}(t_i+t)$ into Eq.(\ref{UU}), we have 
\begin{equation}
\label{TvU}
t_i = \tau \left( e^{U_i/kT}-1 \right).
\end{equation}

Physically, we can consider the decay process during the time interval $t_i$ in Case B as a ``virtual process'' in Case A. This is equal to doing a time transformation $t'= t_i + t$ in Eq.(\ref{UaTime2}). Therefore, Eq.(\ref{UaTime1}) can be rewritten as
\begin{equation}
\label{UaTime3}
U_a(t) = kT ln\left(1 + \frac{t_i+t}{\tau} \right).
\end{equation}
where $t_i$ is defined by Eq.(\ref{TvU}). Substituting Eq.(\ref{TvU}) into Eq.(\ref{UaTime3}), it goes back to Eq.(\ref{UaTime1}). This means that the physical considerations on $t_i$ is consistent in mathematics.

It is now clear that, by introducing the virtual time interval $t_i$, the vortex activation energy of a flux relaxation process starting with a lower current density $j_i$ can be converted into that of a process starting with the critical current density $j_c$. The physical meaning of $t_i$ is: \textit{the virtual time interval during which the vortex activation energy $U_a(t)$ increases from $0$ to the initial value $U_i$}. This will be further discussed later in the study of the time dependence of current density.

\section{Time dependence of current density}

The time dependence of current density $j(t)$ can be obtained by combing the time dependence of the vortex activation energy $U_a(t)$ (see Eq.(\ref{UaTime3})) with the current dependence of the vortex activation energy $U_a(j)$. Various $U_a(j)$ were proposed and each has particular physical meaning.\cite{Feigel'man1,Feigel'man2,Zeldov1,Zeldov2,Anderson,Ma1} Here we used the inverse-power activation energy \cite{Feigel'man1,Feigel'man2} and logarithmic activation energy  \cite{Zeldov1,Zeldov2} to construct the $j(t)$ respectively.

\subsection{Inverse-power activation energy}

In case of random pinning, collective creep theory \cite{Feigel'man1,Feigel'man2} gives a good description to the vortex behavior. This theory gives a vortex activation energy that has an inverse-power dependence on current density, that is,
\begin{equation}
\label{Inverse}
U_a(j) = \frac{U_0}{\mu} \left[ \left(\frac{j_c}{j}\right)^\mu -1 \right],
\end{equation}
where $\mu$ is a critical exponent related the vortex structure in the superconductor. The number $1$ on the right side is introduced to ensure that $U_a(j_c)=0$.

Substituting Eq.(\ref{UaTime3}) into Eq.(\ref{Inverse}), we have the time dependence of the current density
\begin{equation}
\label{InverseJ}
j(t) = j_c\left[1+\frac{\mu kT}{U_0} ln \left(1+\frac{t_i+t}{\tau}\right)\right]^{-1/\mu}.
\end{equation}

Putting $t=0$ in Eq.(\ref{InverseJ}), we obtain the initial current density

\begin{equation}
\label{Ji}
j_i = j_c\left[1+\frac{\mu kT}{U_0} ln \left(1+\frac{t_i}{\tau}\right)\right]^{-1/\mu}.
\end{equation}

Fitting Eq.(\ref{InverseJ}) to experimental data, we can obtain the critical current density $j_c$ (see Fig.1(a)). This means that we can calculate $j_c$ by measuring a flux relaxation process starting with an arbitrary current density $j_i$. This operation is possible because $j_c$ is mainly determined by the vortex pinning potential, which affects each vortex jump (see Eq.(\ref{UaDiff})). It does not matter what current density the flux relaxation process starts with.

Eq.(\ref{InverseJ}) also shows that, by introducing the virtual time interval $t_i$, a flux relaxation process starting with a current density $j_i$ can be converted into a process starting with the critical current density $j_c$. This can be analyzed in a way analogue to that used in Case A and Case B. The process during $t_i$ is a virtual process, introduced for assisting physical understanding and easy calculation.

In Eq.(\ref{TvU}), we expressed the virtual time interval $t_i$ in terms of the initial activation energy $U_i$. But in practice the measurable physical quantity is the initial current density $j_i$. Therefore, it should be more convenient to express $t_i$ in terms of $j_i$. Inverting Eq.(\ref{Ji}), we have
\begin{equation}
\label{TvJInverse}
t_i = \tau \left\{ exp\left\{\frac{U_0}{\mu kT} \left[\left(\frac{j_c}{j_i}\right)^\mu -1\right]\right\} - 1 \right\}.
\end{equation}

Putting $j_i=j_c$ in Eq.(\ref{TvJInverse}), we have $t_i=0$. Therefore, the physical meaning of the virtual time interval $t_i$ can also be explained using the current density: \textit{the time interval during which the current density reduces from the critical value $j_c$ to the initial value $j_i$}.

Fig.1(a) shows the fitting of Eq.(\ref{InverseJ}) to the experimental data from a $Bi_2Sr_2CaCu_2O_{8+x}$ single crystal. The parameter $\mu=0.12$ indicates that the vortices in the sample are approximately ``single vortex''.\cite{Feigel'man1}

\begin{figure}[htb]
\label{figure1}
\begin{center}
\includegraphics[height=0.6\textwidth]{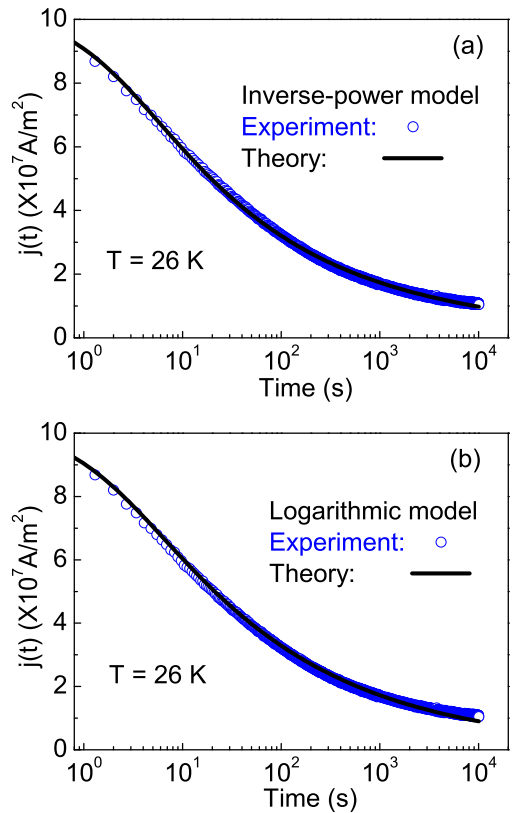}
\caption{(Color online) Time dependence of current density. The scattering points are the persistent current density induced in a $Bi_2Sr_2CaCu_2O_{8+x}$ single crystal at $26$ K. The solid black lines are the theoretical fits.
(a) Fitted with the inverse-power model (Eq.(\ref{InverseJ})).
$j_c=(1.22 \pm 0.01) \times 10^8 A/m^2$, $\mu=(0.12 \pm 0.01)$, $U_0=(79.8 \pm 0.1) k$, $t_i=(0.81\pm0.01)$ s, $\tau=(1.20\pm0.01)$ s.
(b) Fitted with the logarithmic model (Eq.(\ref{LogJ})). 
$j_c=(1.20 \pm 0.00) \times 10^8 A/m^2$, $U_0=(92.2 \pm 0.1) k$, $t_i=(0.79\pm0.01)$ s, $\tau=(1.03\pm0.01)$ s.}
\end{center}
\end{figure}

\subsection{Logarithmic activation energy}

In high-$T_c$ superconductors, sometimes it is more accurate to use the following logarithmic activation energy \cite{Zeldov1,Zeldov2} 
\begin{equation}
\label{Log}
U_a(j) = U_0 ln \left(\frac{j_c}{j}\right).
\end{equation}

Substituting Eq.(\ref{UaTime3}) into Eq.(\ref{Log}), we have the time dependence of the current density
\begin{equation}
\label{LogJ}
j(t) = j_c\left(1+ \frac{t_i+t}{\tau}\right)^{-kT/U_0}.
\end{equation}

Putting $t=0$ in Eq.(\ref{LogJ}), we obtain the initial current density,
\begin{equation}
\label{LogJi}
j_i = j_c\left(1+ \frac{t_i}{\tau}\right)^{-kT/U_0}.
\end{equation}

Inverting Eq.(\ref{LogJi}), we obtain the virtual time interval
\begin{equation}
\label{TvJLog}
t_i = \tau \left[ \left(\frac{j_c}{j_i}\right)^{U_0/kT} - 1 \right].
\end{equation}

Fig.1(b) shows the fitting of Eq.(\ref{LogJ}) to the experimental data from a $Bi_2Sr_2CaCu_2O_{8+x}$ single crystal.

\section{Exact solutions of the time dependence of current density}

We have used Eq.(\ref{UaTime3}) in the derivation of Eq.(\ref{InverseJ}) and Eq.(\ref{LogJ}). Eq.(\ref{UaTime3}) provides a universal method for calculating the time dependence of current density with various vortex activation energies. But the solutions obtained in this way are approximate solutions. Here we show that one can find an exact solution of Eq.(\ref{UaDiff}) using the linear activation energy $U_a(j)=U_0 (1- j/j_c)$ and logarithmic activation energy $ U_a(j) = U_0 ln (j_c/j) $.

\subsection{Linear activation energy}

Substituting the linear activation energy $U_a(j)=U_0 (1- j/j_c)$ into Eq.(\ref{UaDiff}), we have
\begin{equation}
\label{UaDiff2}
\frac{dj}{dt} = - C e^{-\beta(1- j/j_c)},
\end{equation} 
where $\beta=U_0/kT$.

The solution of Eq.(\ref{UaDiff2}) is
\begin{equation}
\label{LinearBb}
j(t) =j_c \left[ 1- \frac{1}{\beta} ln \left( 1+ \frac{t_i+ t}{\tau} \right) \right],
\end{equation}
where $\tau=j_c/(\beta C)$.

Putting $t=0$ in Eq.(\ref{LinearBb}), we obtain the initial current density,
\begin{equation}
\label{LinearJi}
j_i =j_c \left[ 1- \frac{1}{\beta} ln \left( 1+ \frac{t_i}{\tau} \right) \right].
\end{equation}

Inverting Eq.(\ref{LinearJi}), we obtain the virtual time interval
\begin{equation}
\label{TvLinear}
t_i= \tau \left[e^{-\beta(1-j_i/j_c)}-1 \right].
\end{equation}

\subsection{Logarithmic activation energy}

Substituting the logarithmic activation energy $U_a(j) = U_0 ln (j_c/j)$ into Eq.(\ref{UaDiff}), we have
\begin{equation}
\label{UaDiff3}
\frac{dj}{dt} = - C \left(\frac{j}{j_c}\right)^{\beta}.
\end{equation}

The solution of Eq.(\ref{UaDiff3}) is
\begin{equation}
\label{LogBb}
j(t) = j_c \left( 1 + \frac{t_i+ t}{\tau} \right)^{1/(1-\beta)},
\end{equation}
where $\tau= j_c/[(\beta-1)C]$.

Putting $t=0$ in Eq.(\ref{LogBb}), we obtain the initial current density,
\begin{equation}
\label{LogBbJi}
j_i = j_c \left( 1 + \frac{t_i}{\tau} \right)^{1/(1-\beta)}.
\end{equation}

Inverting Eq.(\ref{LogBbJi}), we obtain the virtual time interval
\begin{equation}
\label{TvLog2}
t_i= \tau \left[ \left( \frac{j_i}{j_c} \right)^{(1-\beta)}-1 \right].
\end{equation}

\section{Discussion}

Eq.(\ref{InverseJ}), Eq.(\ref{LogJ}), Eq.(\ref{LinearBb}), and Eq.(\ref{LogBb}) are the time evolution equations of current density with generalized initial conditions. Using these equations, one can calculate the critical current density $j_c$ from a flux relaxation process with a lower initial current density $j_i$.

The inverse-power activation energy Eq.(\ref{Inverse}) is obtained from the collective creep theory \cite{Feigel'man1,Feigel'man2}, which has profound physical meaning. This model also gives the information about the vortex structure by the exponential parameter $\mu$. Therefore, one may choose this model if he/she is interested in finding out the vortex structure in the superconductor.

Sometimes, it is found that the logarithmic activation energy Eq.(\ref{Log}) gives a better description to experimental data.\cite{Zeldov1,Zeldov2} The corresponding time evolution equation of the current density Eq.(\ref{LogJ}) has four free parameters. Therefore, this model is more convenient for analyzing experimental data. One may choose this model if he/she is only interested in finding out the current dependence of the vortex activation energy $U(j)$.

At higher temperatures, however, the flux relaxation phenomena usually exhibit strongly non-logarithmic behavior. The experimental data usually cannot be well fitted by the theoretical models. In this case, one may consider using the infinite series model \cite{Ma1,Ma2}, which gives the information about the elastic and non-elastic deformation of the vortices. This generalized model includes the linear model (see Eq.(\ref{LinearBb})) as a special case. Using this theory, one can calculate the activation energy with high accuracy. But this theory also includes many free parameters and one has to choose the higher order terms on his purpose.

Finally, it should be emphasized that in Ref.~\onlinecite{Ma1} we proposed a generalized infinite series expression for the current dependence of the vortex activation energy, which is related to the difference of the superconductivity in various superconductors. This enables us to analyze the experimental data on the current decay behavior and then calculate the activation energy of a vortex system without subjecting to any special conditions. In the present work, however, we introduced the virtual time interval $t_i$ for the time dependence of a flux relaxation process. This is equal to constructing a generalized initial condition for the time dependence of a flux relaxation process. It is related to the ``time''.

\section{Conclusion}

The information about the critical current density of a type-II superconductor is carried by the vortex system throughout a flux relaxation process. By introducing a virtual time interval, a flux relaxation process starting with a lower current density can be converted into a process starting with the critical current density. This enables us to compute the critical current density from a flux relaxation process with a lower current measurement. It does not matter what initial current density the flux relaxation process starts with.

\end{document}